\documentstyle[preprint,aps]{revtex}

\begin{document}
\title{The Influence of Nonextensivity on Orientational Ordering in Liquid Crystal
Systems with Variable Molecular Shape}
\author{O. Kayacan$\thanks{%
E-mail: ozhan.kayacan@bayar.edu.tr}$}
\address{$^{a}$Department of Physics, Faculty of Arts and Sciences, Celal Bayar\\
University, Muradiye, Manisa-TURKEY}
\maketitle

\begin{abstract}
A generalized model taking into account the photoisomerization influence on
the nematic ordering is presented. This generalized theory is used to
investigate the effect of the nonextensivity on the concentration dependence
of the long-range order parameter. The $q$-dependent variation of the
concentration of $cis-trans$ \ isomers and of the order parameter with the
time of exposure to the illumination is investigated within nonextensivity.
It is also shown that for sufficiently long exposition of the mesophase to
the illumination the nematic phase cannot disappear for some value of the
entropic index, coming from Tsallis statistics. Moreover it is shown that
long range interactions or the fractal structure in the liquid crystalline
system might affect the characteristics of the phase transition in the
physical system. We think therefore that this conclusion might shed light on
the interaction potential energy terms in the similar system in future
possible experiments, even to be performed on new objects with different
symmetries.

\end{abstract}

\newpage

\section{Introduction}

A nonextensive entropy definition was proposed in 1988 [1] by Tsallis, 
\begin{equation}
S_{q}=-k\frac{1-\sum \,_{i=1}^{W}\;p_{i}^{q}}{1-q},\;\;q\epsilon R
\end{equation}
where $k$ is a positive constant, {\it p}$_{\text{{\it i}}}$\ is the
probability of the system in the {\it i} th microstate, {\it W}\ is the
total number of the configurations of the system and {\it q }is any real
number. Since then, Tsallis thermostatistics (TT) has commonly been used for
the investigation of the physical systems. As well-known, Boltzmann-Gibbs
(BG) statistics is a powerful one to study a variety of the physical
systems. However it falls for the systems which $1)$ have long-ranged
interactions, $2)$ have long-ranged memory effects, and $3)$ evolve in
multifractal space-time. In this manner, it has been understood that
extensive BG statistics fails to study nonextensive physical systems not
having these conditions. Consequently the standard BG thermostatistics is
not universal and is appropriate for extensive systems. The parameter, $q$,
in Eq.(1) is called the entropy index and measures the degree of the
nonextensivity of the system under consideration. After a few years from
1988, the formalism was revised [2] introducing the unnormalized constraint
to the internal energy. This generalization included the celebrated BG
thermostatistics, which could be recovered when $q=1$. After this work, this
formalism has been commonly employed to study various physical systems. Some
of them can be given such as; self gravitating systems [3,4], turbulence
[5-8], anomalous diffusion [9-12], solar neutrinos [13], liquid crystals
[14-18] etc.

In spite of the successes of TT, some drawbacks were noted in the original
formalism. These can be summarized as follows: $i)$ the density operator was
not invariant under a uniform translation of energy spectrum, $ii)$ the $q$%
-expected value of the identity operator was not the unity, $iii)$ the
energy of the physical system was not conserved. To overcome these
difficulties, Tsallis et al. introduced the normalized constraint to the
internal energy of the system [19]: 
\begin{equation}
\frac{\sum_{i=1}^{W}p_{i}^{q}\varepsilon _{i}}{\sum_{i=1}^{W}p_{i}^{q}}%
=U_{q}{}.
\end{equation}
The optimization of Tsallis entropy given by Eq.(1) according to this
internal energy constraint results in 
\begin{equation}
p_{i}=\frac{\left[ 1-(1-q)\beta ^{\ast }(\varepsilon
_{i}-U_{q})/\sum_{j=1}^{W}(p_{j})^{q}\right] ^{\frac{1}{1-q}}}{Z_{q}}
\end{equation}
with 
\begin{equation}
Z_{q}=\sum_{i=1}^{W}\left[ 1-(1-q)\beta ^{\ast }(\varepsilon
_{i}-U_{q})/\sum_{j=1}^{W}(p_{j})^{q}\right] ^{\frac{1}{1-q}},
\end{equation}
where 
\[
\beta ^{\ast }=\frac{\beta }{\left( 1/Tr\left\{ \widehat{\rho }^{q}\right\}
\right) +\left( 1-q\right) U_{q}\beta },\;\;\beta =1/kT. 
\]

Therefore the $q$-expectation value of any observable is defined as 
\begin{equation}
\left\langle A_{i}\right\rangle _{q}=\frac{\sum_{i=1}^{W}p_{i}^{q}A_{i}}{%
\sum_{i=1}^{W}p_{i}^{q}},
\end{equation}
where $A$ represents any observable quantity which commutes with
Hamiltonian. As can be expected, the $q$-expectation value of the observable
reduces to the conventional one when $q=1$.

As for our motivation in this study, it was found in literature [20] that
the structure of the micellar and liquid crystalline phases can be
considered as self-similar. Based on this fact, recently a model of
self-similar aggregation of alkylbenzenesulphonates (ABS) molecules was
proposed [21]. Then the fractal structure of the sample was investigated
experimentally [20] where the influence of the geometry of ABS molecules on
the formation of their liquid crystalline phases was reported and the
authors emphasized that ABS molecules are the fractal objects. As mentioned
above, the nonextensivity may appear in systems where fractality exists.
Therefore the main motivation of the present study comes from having fractal
structure of some liquid crystals.

As well-known from the literature and our last studies on nematic liquid
crystals, Maier-Saupe theory (MST), which is a mean-field one, has some
insufficiencies coming from; 1) as being a mean-field theory, 2) neglecting
higher interaction terms in the interaction potential function. Due to these
insufficiencies, MST gives, for instance, the critical value of the long
range order parameter $\overline{P_{2}}=0.429$ for all nematics. Considering
these terms (i.e. the possibility of the fractality in liquid crystals and
the insufficiencies of the mean-field approach), we apply TT to the nematic
liquid crystals. To this end, in [14], MST was generalized within TT and it
was showed the influences of the nonextensivity on the nematic-isotropic
transition. Next the dimerization process was handled using TT in [15] which
can be regarded as the first application of TT to a liquid crystalline
system. In [15], the influence of the dimerization process on the nematic
ordering is investigated by using a nonextensive thermostatistics (TT). A
theoretical model taking into account the dimerization influence on the
nematic scalar order parameter has been summarized and theoretical
predictions for the nematic order parameter are improved by using TT. Then
the $P_{4}$ model with $P_{4}$-interaction ($P_{4}$ indicates the Legendre
polynomial for $L=4$) was investigated and generalized within TT in [16].
The dependence of the second order parameter ($\overline{P_{2}}$) on the
fourth one ($\overline{P_{4}}$) was studied\ \ in [17] and the generalized
theory was applied to PAA (para-azoxyanisole), a nematic liquid crystal. In
[18], a generalized mean-field theory was presented relating helix tilt in a
bilayer to lipid disorder. It has been observed from all these studies that
if one uses a generalized form of the MST within TT, the deviations between
the theory (MST) and the experimental data are minimized and the obtained
results agree with the experimantal data very well.

In this study, the influence of the photoisomerization on the orientational
ordering is handled by using a generalized theory. The standard theory has
been recently proposed in literature [22]. It is an extension of the MST and
could be an appropriate one to investigate the variation of the
concentration of $cis-trans$ isomers and of the scalar order parameter with
the time of exposure to the illumination. As expected, as it is an extension
of a mean-field theory (MST), it might have some deviations form the
experimental data; for instance it also gives a universal value of the long
range order parameter at transition point, $0.429$. To overcome similar
insufficiencies (considering the possible fractality of the liquid
crystalline system), we use a generalized form of the standard theory and
investigate the physical quantities concerning physical system.\ To this
end, we first summarize the standard and generalized theories. Then we
investigate the effects of nonextensivity on the nematic-isotropic
transition and also on the evolution of the concentration of $cis-trans$
isomers and of the scalar order parameter with the time of exposure to the
illumination.

\section{The Generalized Theory}

The standard theory [22] assumes that the liquid crystalline system is
formed by a mixture of two elements, namely two miscible nematogens $N_{1}$
and $N_{2}$. According to the system under consideration can be dimers and
monomers, if an energy to break the hydrogen bonding is considered [23], or
trans and cis isomers in the case of azobenzene materials [24]. Both kinds
of molecules, namely $N_{1}$ and $N_{2}$, are supposed to be rodlike. In the
mean-field approximation, the total nematic potential function can be given
by [25-27] 
\begin{equation}
V\left( \theta _{1},\theta _{2}\right)
=\sum_{i,j}V_{ij}=V_{11}+V_{12}+V_{21}+V_{22},
\end{equation}
where $\theta _{1}$ and $\theta _{2}$ are the angles formed by the molecular
long axes of the molecules of $N_{1}$ and $N_{2}$ with the nematic director
of the mixture respectively, $V_{11}$ and $V_{12}$ are the mean-field
potentials acting on a molcule of $N_{1}$ due to the other molecules of $%
N_{1}$ and due to the molecules of $N_{2}$ respectively. A similar meaning
has $V_{21}$ and $V_{22}$. These potential functions can be written as 
\begin{equation}
V_{ij}=-\alpha _{ij}P_{2}\left( \theta _{i}\right) \left\langle P_{2}\left(
\theta _{j}\right) \right\rangle ,
\end{equation}
in the MST, where $\alpha _{ij}$'s are coupling constants and they depend on
the distance between the center of mass of the molecules, on a molecular
property and also on the concentration of one of the constituents in the
system [23]. Using these terms above, we can write the nematic potential
function as 
\begin{eqnarray}
V\left( \theta _{1},\theta _{2}\right) &=&V_{1}(\theta _{1})+V_{2}(\theta
_{2})  \nonumber \\
&=&-\left( \alpha _{ii}\overline{P_{2}(\theta _{i})}+\alpha _{ij}\overline{%
P_{2}(\theta _{j})}\right) \,P_{2}(\theta _{i}),
\end{eqnarray}
where $\overline{P_{2}(\theta _{i})}=\left\langle P_{2}(\theta
_{i})\right\rangle $ is the scalar order parameters of the molecules of $%
N_{i}$.\ The scalar order parameters are determined by solving the set of
coupled equations according to Eq.(5) for $i=1,2$ self consistently. These
coupled equations depend on the temperature $T$ and on the quantities $%
\alpha _{ij}$ appearing in the nematic potential function. The $\alpha _{ij}$%
\ quantities contain the strengths of mean-field interaction among the
molecules forming the nematic $1$, among the molecules forming the nematic $%
2 $, and among the molecules of $1$ and those of $2$. $\alpha _{ij}^{\prime
}s$ are assumed as 
\begin{equation}
\alpha _{ij}=u_{ij}n_{j},
\end{equation}
where $u_{ij}=u_{ji}$ and $n_{i}=N_{i}/(N_{1}+N_{2})$, with $N_{i}$ being
the number of molecules of the nematic $i$ [23].

The generalized free energy of the system is given by 
\begin{equation}
F_{q}=-\sum_{i}\left( \frac{1}{2}N_{i}\left\langle V_{i}\right\rangle
+N_{i}kT\;\ln _{q}(Z_{i,q})\right) ,
\end{equation}
where $ln_{q}(Z_{i,q})=(Z_{i,q}^{1-q}-1)/(1-q)$ and $\left\langle
V_{i}\right\rangle =-\sum_{j}\alpha _{ij}P_{2}\left( \theta _{j}\right)
P_{2}\left( \theta _{i}\right) $. Since we assume $u_{12}=u_{21}=u$, the
free energy density of the system can be given by 
\begin{equation}
F_{q}=N\left\{ -\sum_{i}\left[ \frac{1}{2}n_{i}^{2}u_{ii}\left( P_{2}\left(
\theta _{i}\right) \right) ^{2}+n_{i}kT\ln _{q}(Z_{i})\right]
+un_{1}n_{2}P_{2}\left( \theta _{1}\right) P_{2}\left( \theta _{2}\right)
\right\} ,
\end{equation}
where $N=N_{1}+N_{2}$.

Let us now consider the case in which one of the constituents do not
contribute to the nematic order. The experimental evidence shows that the
molecules of the cis isomers are of banana shape. According to\ recent
investigations, this banana shape can be approximated by two linear parts
forming an angle between them not very far from $\pi /2$ [22]. As a result
of this configuration, the molecules of the cis isomer behave as spherical
objects due to the spinning and tumbling motions of thermal origin.
Consequently in the liquid phase they can originate just an isotropic phase.
In this manner, the molecules of cis isomer lead to a kind of neutral
background, where the molecules of trans isomer are responsible for nematic
order.\ So we will consider that the trans isomers have cylindrical symmetry
and can exhibit a nematic order, whereas the cis isomers will tend to reduce
this orientational order. In addition, we also assume that at zero
temperature the system under consideration is made by $N_{trans}=N$ isomers.
At a given temperature $T$ we have 
\begin{equation}
n_{trans}+n_{cis}=1,
\end{equation}
where $n_{trans}$ and $n_{cis}$ are the equilibrium concentration of trans
and cis isomers and are given by 
\begin{eqnarray*}
n_{trans} &=&N_{trans}(T)/N \\
n_{cis} &=&N_{cis}(T)/N
\end{eqnarray*}
at a given temperature respectively. Thus the quantities $\alpha _{ij}$ are
reduced to the case of trans isomer-trans isomer interaction, i.e., $\alpha
_{ij}=\alpha _{trans}$ and $\alpha _{trans}=un_{trans}$ [23].

In the case of one component, $\overline{P_{2}(\theta _{i})}=\left\langle
P_{2}(\theta )\right\rangle $\ will be the scalar order parameter of trans
isomers. The standard theory is based on MST with a ingredient represented
by temperature-dependent concentration. This dependence in the standard
theory plays an important role and it departs from the one represented by
the concentration dependence characteristic of the excluded volume theories
of Onsager type [28,29]. As well-known, in Onsager model the number of
particle is fixed. In this manner, the critical density, below which the
isotropic phase is stable, follows from a balance between the energy
connected with the excluded volume and the thermal energy. In contrast,
according to the standard model which will be generalized below, the number
of particles responsible for the nematic phase depends on the temperature.
In Onsager's theory, this is equivalent to having an excluded volume that is
temperature dependent.

In Fig.(1), we show the dependence of the long range order parameter on the
concentration of cis isomers for a fixed reduced temperature $T_{R}=0.1\,u/k$
and for some $q$ values. The nematic scalar order parameter vs. the
concentration of the cis isomers in the system for $T_{R}=0.1\,u/k$ (reduced
temperature) is illustrated in Fig.(1). Since the standard theory ($q=1$)
depends on Maier-Saupe mean-field interaction energy, the system, as
expected, exhibits a first-order phase transition to the isotropic phase for 
$n_{cis}^{c}=0.55$ and $\left( \overline{P_{2}}\right) _{c}$ decreases
suddenly to zero from $0.429$ which MST gives for all nematic liquid
crystals. However it is experimentally known [30] that the critical value of
the nematic order parameter varies from one nematic to the other and it lies
in a range of $0.25-0.5$ for nematic liquid crystals. We think therefore
that the standard model based on mean-field approach appears to be
insufficient and alternative approach is needed. On the other hand, as $q$
is varied, we observe that the critical value of the order parameter
changes. The obtained results are summarized in Table in which the entropy
change at transition point is also reported as a function of $q$. We denote
the entropy of the system as $S_{q}$.

In Fig.(2), the phase diagram of temperature vs. concentration of cis
component for the system is plotted. One observes that the usual critical
temperature ($T_{R}\simeq 0.2202u/k$) for MST [31] is reached when $%
n_{cis}=0 $ as expected, since the concentration of the molecules leading to
the nematic phase is fixed in MST. An interesting point is that the
experimental data reported in [25] show that the critical temperature for
the nematic-isotropic phase transition of the mixture is a linear function
of the concentration of the solute. In this framework, we observe from
Fig.(2) that the entropic index $q$ just affects the slope of the line
associated with this linear dependency.

\section{Order Parameter vs. Exposure Time for illumination}

At a given temperature, the isomerization reaction follows the reaction
scheme $trans\rightarrow cis$, depending on some rate constants that, in
turn, depend on the intensity of the light in the illumination process
[32,33]. Then time variation of the concentrations of two constituents can
be approximated by 
\begin{equation}
n_{trans}(T,t)=c\,n_{trans}(T)+(1-c)\,n_{trans}(T)\,\exp _{q}(-t/\tau )
\end{equation}
within TT, where $\tau $ is a characteristic time and $c$ is a parameter to
control the fraction of trans-cis isomers at a given temperature after the
illumination. When $t=0$, we have $n_{trans}(T,0)=n_{trans}(T)$. From $%
n_{trans}+n_{cis}=1$ and assuming for $n_{trans}$ the approximate expression
given by Eq.(13), we obtain 
\begin{equation}
n_{cis}(T,t)=1-n_{trans}(T,t).
\end{equation}
It is the experimental fact that with the exposure time to the illumination,
the concentration of cis isomers increases, whereas the trans isomer
concentration does not tend to zero, but to some fraction, controlled by the
parameter $c$, of the equilibrium concentration $n_{trans}(T)$.

Now we would like to discuss about the equilibrium distribution of the
concentrations of the different constituents forming the mixture. At a given
temperature, the equilibrium distributions of cis and trans isomers in TT
could be given by 
\begin{eqnarray}
n_{cis} &=&\exp _{q}(-\beta (\mu +E))  \nonumber \\
n_{trans} &=&\exp _{q}(-\beta \mu )
\end{eqnarray}
respectively, where $\mu $ is the chemical potential of the mixture. Now if
the above results are used in Eq.(12), it is possible to observe the
behaviour of the order parameter with the exposure time to the illumination.
In Fig.(3), we illustrate the variation of the order parameter with the
exposure time for the reduced temperature of $T_{R}=0.1u/k$, $c=0.3$ and for
some $q$ values. We observe that for a given value of c, the entropic index $%
q$ plays an important role and has a considerable effect on the variation of
scalar order parameter with the exposure time. The similar behaviour can be
easily seen in Fig.(4) in which the order parameter vs. exposure time is
plotted for $c=0.4$ and some values of $q$. Probably the most interesting
result is seen in Fig.(5); for $q=1$ (standard theory) and $c\geq 0.55$,
there is no phase transition at temperature $T_{R}=0.1u/k$. However for $%
q\leq 0.5$ and $c=0.55$, a first order phase transition occurs at this
temperature ($T_{R}=0.1u/k$). This conclusion implies that the permenance of
the liquid crystalline phase depends not only on the concentration of cis
isomers in the system, but also on the entropic index $q$ (i.e. probably
long range interactions or fractality).

\[
\begin{tabular}{|l|l|l|l|l|l|}
\hline
$q$ & $\ 0.98$ & $\ 0.99$ & $\ \ \ 1$ & $\ 1.01$ & $1.02$ \\ \hline
$s_{c}$ & $0.346$ & $\ \ 0.391$ & $0.429$ & $0.463$ & $\ 0.49$ \\ \hline
$\Delta S_{q}/Nk$ & $-0.281$ & $-0.353$ & $-0.418$ & $-0.479$ & $-0.528$ \\ 
\hline
\end{tabular}
\]

\qquad \qquad \qquad \qquad \qquad \qquad \qquad \qquad Table.

\section{Summing up}

We have presented a generalized model to study on the behaviour of the
scalar order parameter in liquid crystalline systems, consisting of more
than one nematogen component. It is the experimental fact that some liquid
crystals have fractal structure. This fact implies that TT can be applied to
the liquid crystalline systems. In addition,\ the standard model, which has
been generalized, is an extension of MST and thus some drawbacks coming from
insufficiencies of mean-field approach could be expected. In this framework,
a generalized model which is an extension of MST has been used. The
generalized theory\ can be successfully applied to distinct systems, such as
one formed by cyclic dimers and monomers, and also to the systems where
azobenzene compounds play an important role. For example, a similar system,
consisting of closed and open dimers, has been studied in [15], where the
experimental data related to the dimeric system has been successfully
explained by using a generalized theory. It must be noted that the success
of the generalized mean-field theories in our earlier studies may come from
being fractal structure of liquid crystals, apart from considering long
range interactions. Similar successful explanations can be expected in other
dimeric and liquid crystalline systems.

It has been also studied on the systems where the illumination process gives
rise to the appearance of new objects with different symmetries in the
systems. We have shown that fractality or long range interactions (i.e. $%
q\neq 1$) might affect the characteristics of the phase transition in the
physical system. This conclusion might shed light on the interaction
potential energy terms or fractal structure in the similar system in future
possible experiments, even to be performed on new objects with different
symmetries.

\newpage

\section*{REFERENCES}

[1] C. Tsallis, J. Stat. Phys. 52 (1988) 479.

[2] E. Curado, C. Tsallis, J. Phys. A: Math. Gen. 24 (1991) L69.

[3] A.R. Plastino, A. Plastino, Phys. Lett. A 174 (1993) 384.

[4] J.J. Aly, Proceedings of N-Body Problems and Gravitational Dynamics
(Publications de L'Observatoire de Paris, Paris, 1993), p. 19, eds. F.
Combes and E. Athanassoula.

[5] T. Arimitsu, N. Arimitsu, Phys. Rev. E 61 (2000) 3237.

[6] T. Arimitsu, N. Arimitsu, J. Phys. A 33 (2000) L235.

[7] C. Beck, Physica A 277 (2000) 115.

[8] C. Beck, G. Lewis, H. Swinney, Phys. Rev. E 63 (2001) 035303.

[9] A. Plastino, A. Plastino, Physica A 222 (1995) 347.

[10] C. Tsallis, D. Bukman, Phys. Rev. E 54 (1996) R2197.

[11] M. Bologna, C. Tsallis, P. Grigolini, Phys. Rev. E 62 (2000) 2213.

[12] L. Malacarne, L. Pedron, R. Mendes, E. Lenzi, Phys. Rev. E 63 (2001)
30101R.

[13] A. Lavagno, P. Quarati, Nucl. Phys. B 87 (2000) 209.

[14] O. Kayacan, F. B\"{u}y\"{u}kk\i l\i \c{c}, D. Demirhan, Physica A 301
(2001) 255.

[15] O. Kayacan, Physica A 325 (2003) 205.

[16] O. Kayacan, Chemical Physics 297 (2004) 1.

[17] O. Kayacan, Physica A 337 (2004) 123.

[18] O. Kayacan, Biophys. Chem. 111 (2004) 191.

[19] C. Tsallis, R. Mendes, A. Plastino, Physica A 261 (1998) 534.

[20] D. Tezak, N. Jalsenjak, S. Puncec, M. Martinis, Colloids and Surfaces A
128 (1997) 273.

[21] D. Tezak, M. Martinis, S. Puncec, I. Fischer-Palkovic, F. Strajnar,
Liq. Crystals 19 (1995) 159.

[22] G. Barbero, L.R. Evangelista, Phys. Rev. E 61 (2000) 2749.

[23] G. Barbero, L.R. Evangelista, M.P. Petrov, Phys. Lett. A 256 (1999) 399.

[24] K. Ichimura, in Photochromism: Molecules and Systems, edited by H.
D\"{u}rr and H. Bouas-Laurent (Elsevier, Amsterdam, 1990).

[25] D.E. Martire, in The Molecular Physics of Liquid Crystals, edited by
G.R. Luckhurst and G.W. Gray (Academic Press, Londn, 1974).

[26] P.Palffy-Muhoray, D.A. Dunmur, W.H. Miller, D.A. Balzarini, in Liquid
Crystals and Ordered Fluids, edited by Angelm C. Giffin and Julien F.
Johnson (Plenum, New York, 1984), Vol. 4.

[27] S. Grande, A. Kuhnel, F. Seifert, Liq. Cryst. 4 (1989) 625.

[28] L. Onsager, Ann. (N.Y.) Acad. Sci. 51 (1949) 627.

[29] R. Zwanzig, J. Chem. Phys. 39 (1963) 1714.

[30] A. Beguin, J.C. Dubois, P. Le Barny, J. Billard, F. Bonamy, J.M.
Busisine, P. Cuvelier, Mol. Cryst. Liq. Cryst. 115 (1984) 1.

[31] E.B. Priestley, P.J. Wojtowicz, P. Sheng, Introduction to Liquid
Crystals (Plenum Press, New York, 1976).

[32] H. Bach, K. Anderle, T. Fuhrmann, J.H. Wendorf, J. Phys. Chem. 100
(1996) 4135.

[33] Y. Ichimura, J. Akita, H. Akiyama, K. Kudo, Y. Hayashi, Macromolecules
30 (1997) 903.

\newpage

\section*{\bf FIGURE AND TABLE CAPTIONS}

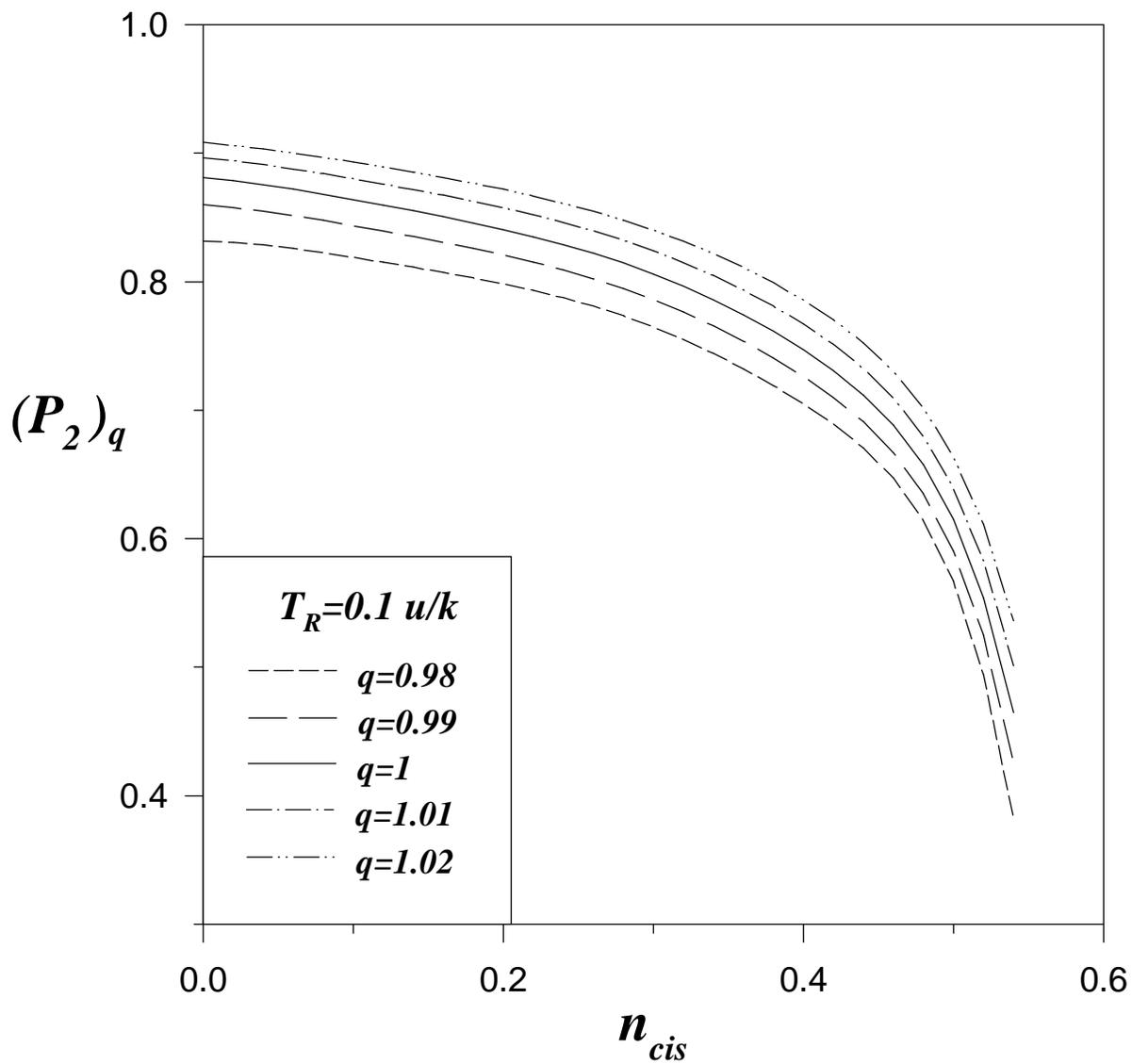
Figure 1. The variation of the scalar order parameter with the concentration
of cis isomer for a fixed reduced temperature $T_{R}=0.1u/k$ and for some $q$
values. As $q$ increases, the critical value of the scalar order parameter
at transition point also increases.

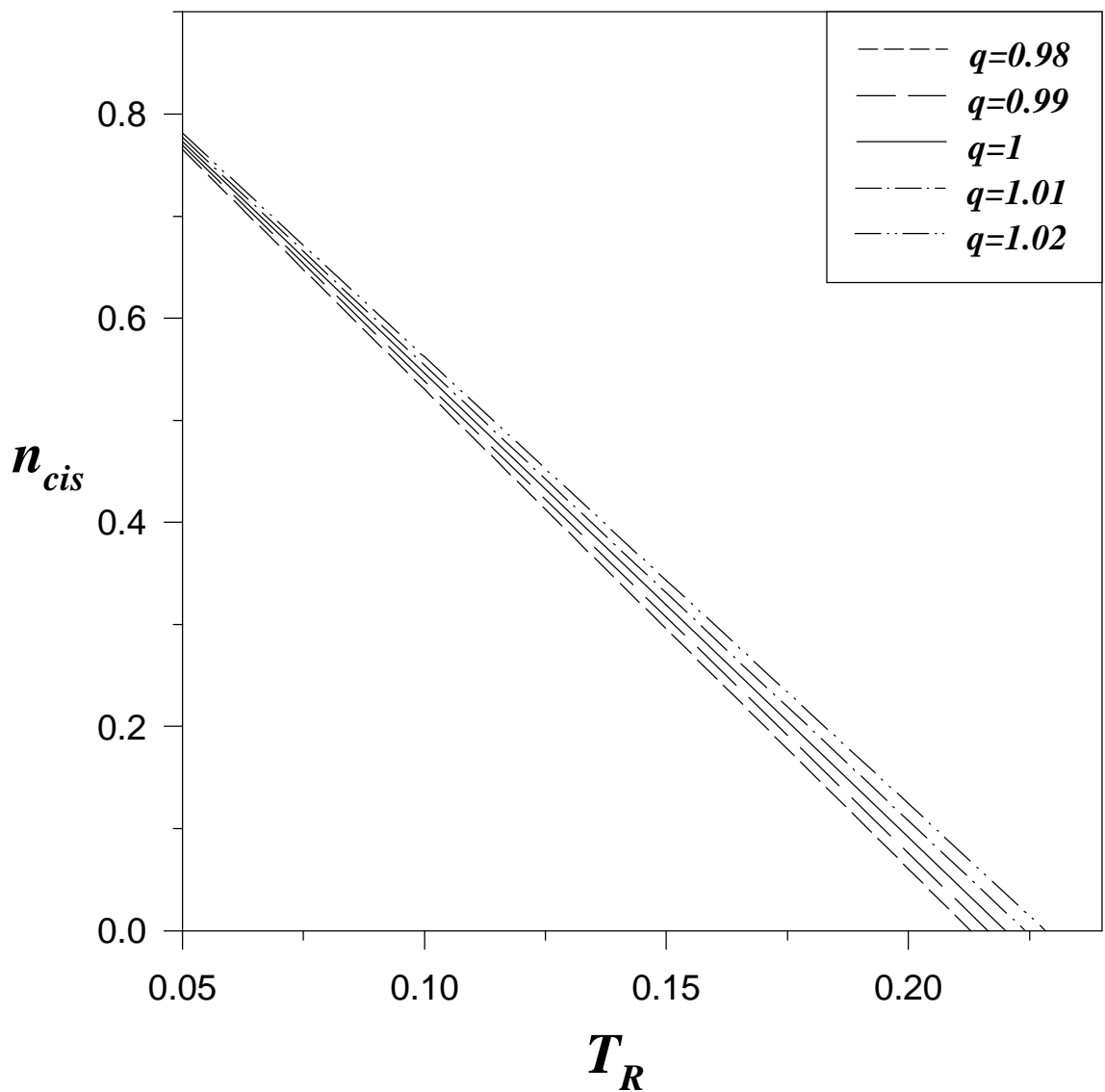
Figure 2. $q$-dependent phase diagram in the reduced temperature vs. cis
concentration plane. The critical lines represent the first order phase
transition, namely nematic-isotropic transition, induced by the variation in
the concentration of cis isomers in the nematic medium.

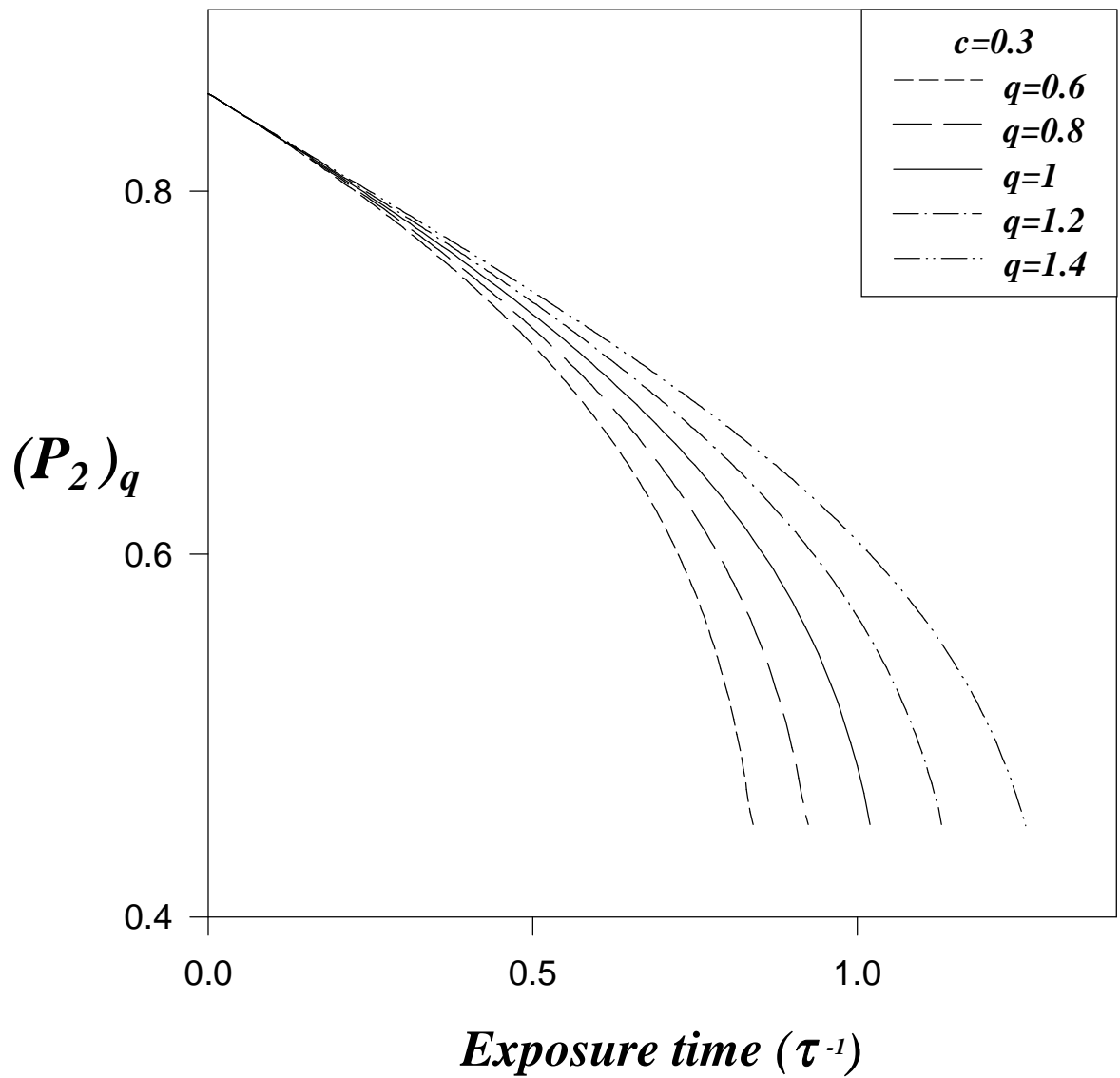
Figure 3. The variation of the scalar order parameter with the exposure time
for uv illumination for $c=0.3$ and some $q$ values. $\tau $ is a
characteristic time. As seen from figure, the entropic index $q$ affects the
variation of scalar order parameter with the exposure time considerably.

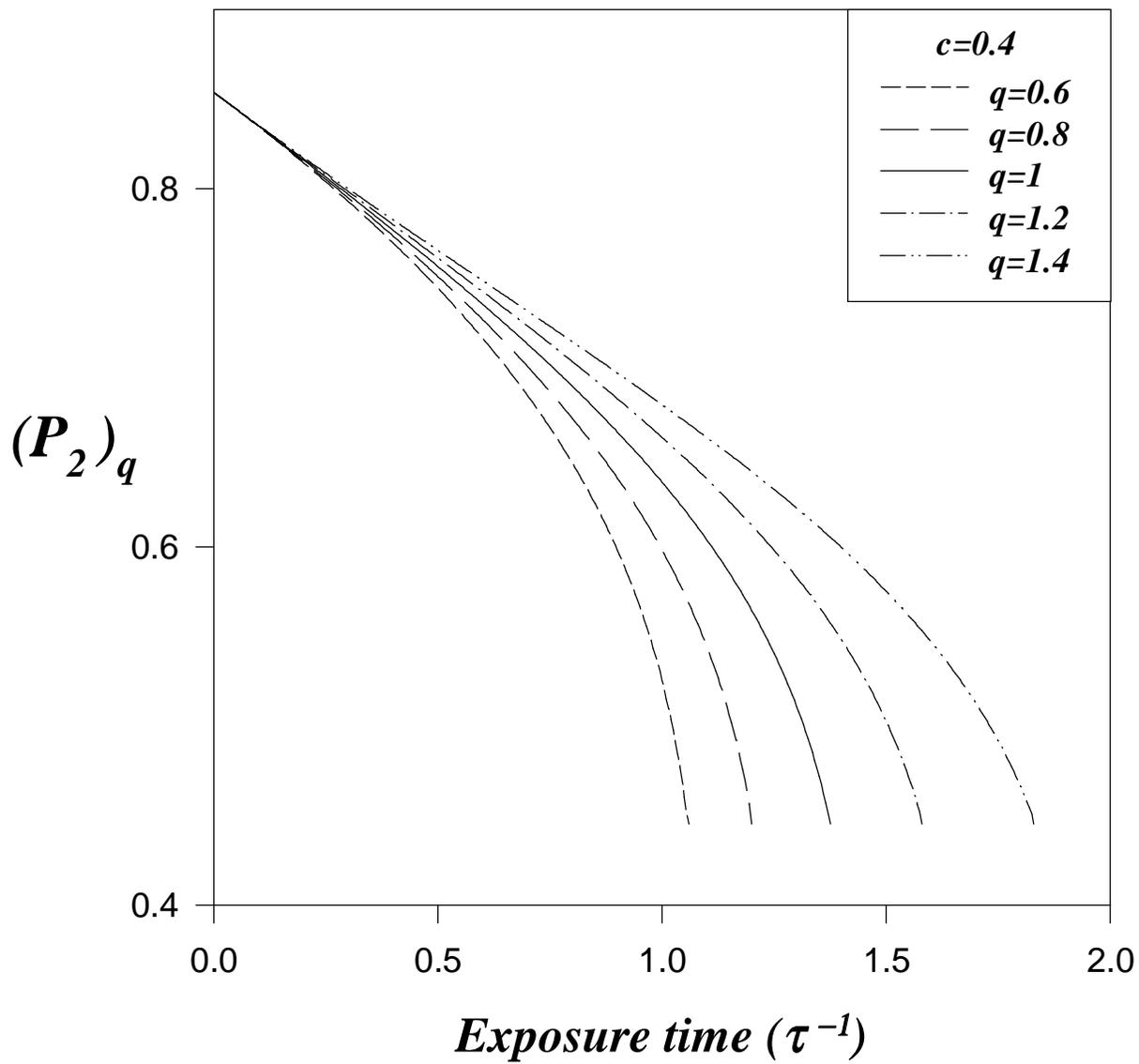
Figure 4. The variation of the scalar order parameter with the exposure time
for uv illumination for $c=0.4$ and some $q$ values. As seen from figure,
the entropic index $q$ affects the variation of scalar order parameter with
the exposure time considerably.

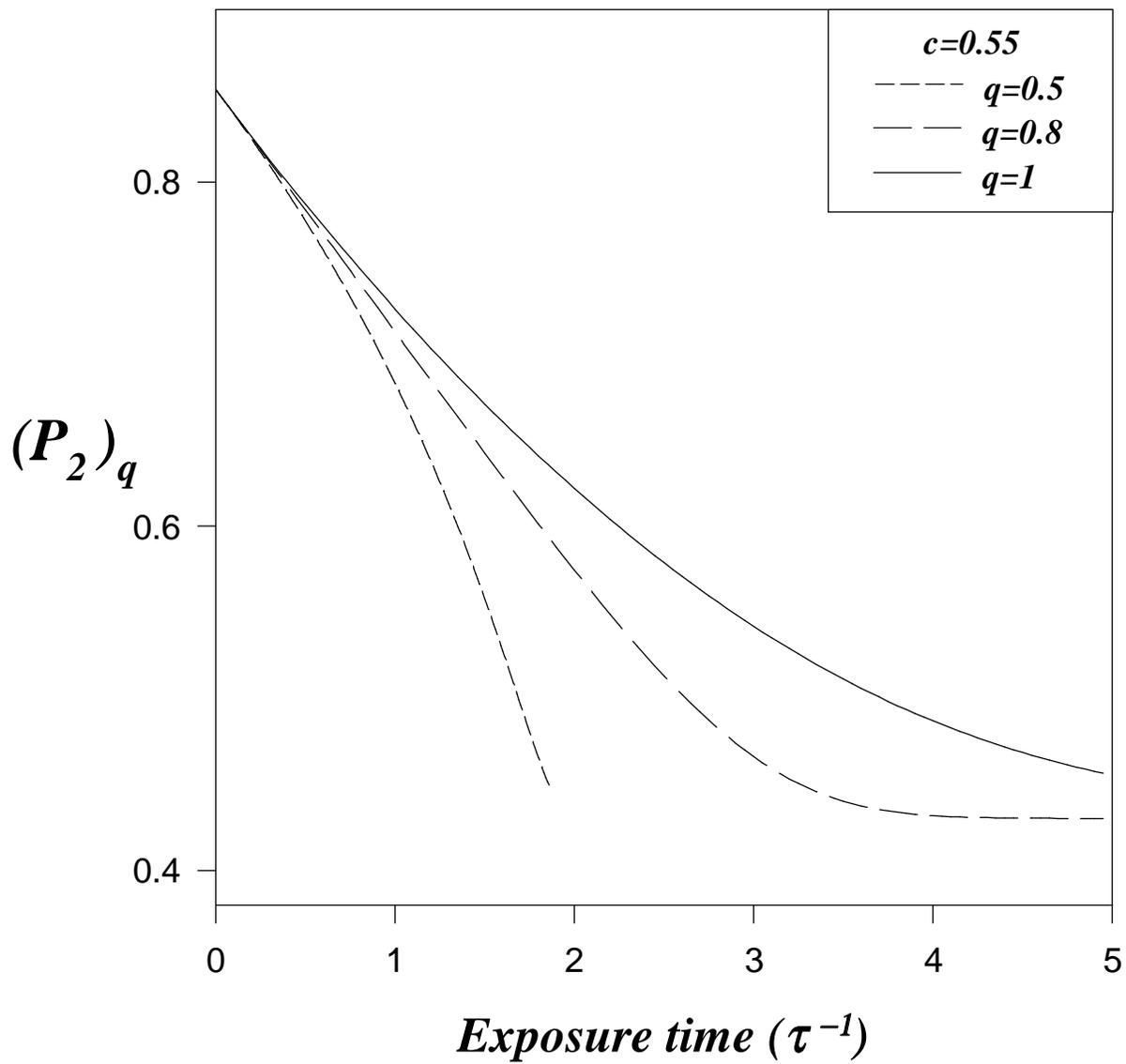
Figure 5. The variation of the scalar order parameter with the exposure time
for uv illumination for $c=0.55$ and some $q$ values. As seen from figure,
whereas there is no phase transition in the standard theory ($q=1$) for $%
c\geq 0.55$, the generalized theory assumes a first order phase transition
for $q\leq 0.5$.

Table. Entropy change of the system at transition point for some values of
the entropic index $q$.

\noindent

\end{document}